\documentclass{article}
\usepackage{authblk}
\usepackage{graphicx}
\usepackage{color}
\newcommand{\be}{\begin{equation}}
\newcommand{\ee}{\end{equation}}
\usepackage{amssymb,amsfonts,amsmath}
\usepackage{geometry}
\begin{document}

\title{Single-cell analysis of growth in budding yeast and bacteria reveals a common size regulation strategy}
\date{}
\author[1]{Ilya Soifer}
\author[2,3,4]{Lydia Robert}
\author[5]{Ariel Amir}
\affil[1]{Department of Molecular Genetics, Weizmann Institute of Science, Rehovot 76100, Israel}
\affil[2]{INRA, UMR1319 Micalis, F-78350 Jouy-en-Josas, France}
\affil[3]{AgroParisTech, UMR Micalis, F-78350 Jouy-en-Josas, France}
\affil[4]{Laboratoire Jean Perrin UPMC-CNRS UMR8237, UPMC, 75005 Paris, France}
\affil[5]{School of Engineering and Applies Sciences and Department of Physics, Harvard University, Cambridge, Massachusetts 02138, USA, arielamir@seas.harvard.edu}

\maketitle

\vspace{-1cm}
\section*{Summary}
To maintain a constant cell size, dividing cells have to coordinate cell cycle events with cell growth. This coordination has for long been supposed to rely on the existence of size thresholds determining cell cycle progression \cite{kafri}. In budding yeast, size is controlled at the G1/S transition \cite{review2}. In agreement with this hypothesis, the size at birth influences the time spent in G1: smaller cells have a longer G1 period \cite{johnston1977coordination}. Nevertheless, even though cells born smaller have a longer G1, the compensation is imperfect and they still bud at smaller cell sizes. In bacteria, several recent studies have shown that the incremental model of size control, in which size is controlled by addition of a constant volume (in contrast to a size threshold), is able to quantitatively explain the experimental data on 4 different bacterial species \cite{amir_PRL_size, sattar, jw, xavier}. Here, we report on experimental results for the budding yeast \emph{Saccharomyces cerevisiae}, finding, surprisingly, that cell size control in this organism is very well described by the incremental model, suggesting a common strategy for cell size control with bacteria. Additionally, we argue that for \emph{S. cerevisiae} the ``volume increment" is not added from birth to division, but rather between two budding events.

\section*{Results and discussion}
\subsection* {Correlations between cell cycle variables support the incremental model}

Cells of all kingdoms of life have to coordinate cell cycle events and cell growth \cite{kafri}. Here, we study cell size control in the budding yeast \emph{S. cerevisiae}, focusing on the diploid form prevalent in nature, and show that it bears much similarity with size control in \emph{Escherichia coli}. To that end, we devised an imaging system that allows following the unperturbed cell cycle dynamics of thousands of yeast cells, identifying sizes and various events of cell cycle with high temporal resolution (Figs. S1A-B). Experiments were performed with cells growing in 5 different culture media. We focus on daughter cells (the first cell cycle of each cell after budding off mother cell, see Fig. 1A), which are known to have stronger size control than mother cells \cite{review2}.
Sizes at birth and division of daughter cells were strongly correlated in all conditions (Fig. 1C). Upon averaging the data for all cells with a given size at birth (suppressing the effects of biological stochasticity), we found for all 5 growth conditions a linear correlation with a slope very close to 1 (slopes within 10\% of 1 in all 5 growth conditions, see caption of Fig. 1C) i.e.:

\be v_d= v_b+\Delta, \label{incremental}
\ee
where $v_b$ is the cell size at birth, and $v_d$ at division. See further details of the data analysis procedure in the SI, and Figs. S1C-E. The slope of 1 under very different growth conditions is evidence to the robustness of our results, and is supportive of the incremental model, recently shown to be the size control mechanism for 4 different species of bacteria \cite{amir_PRL_size, sattar, jw, xavier} -- but has not been shown before to be applicable to budding yeast. In this model a constant volume is effectively added from birth to division, as described by Eq. (\ref{incremental}). To further test the incremental model, we considered the \emph{time} between cell birth and division, $t_d$. In the SI we show that growth at the single-cell level is exponential (Fig. S2A, see also Refs. \cite{jun, godin, amir_bending, sri2, size2, siggia}). Therefore, the time needed in order to add a constant $\Delta$ between birth and division is given by:

\be t_d = \frac{1}{\lambda} \log(1+\Delta/v_b), \label{time_corr} \ee
where $\lambda$ is the growth rate. This implies a specific correlation between interdivision time and size at birth, where all parameters can be independently determined: the growth rate in each medium is experimentally extracted independently, and the constant $\Delta$ is extracted from the size-size correlations of Eq. (\ref{incremental}) and Fig. \ref{size_size}. The excellent agreement of this prediction with the data is shown in Fig. 2, with \emph{no fitting parameters}.

\subsection*{The incremental model is implemented at the G1/S transition}
\label{budding}

\begin{figure}[h]
\vspace{-3cm}
\includegraphics[width=1 \linewidth]{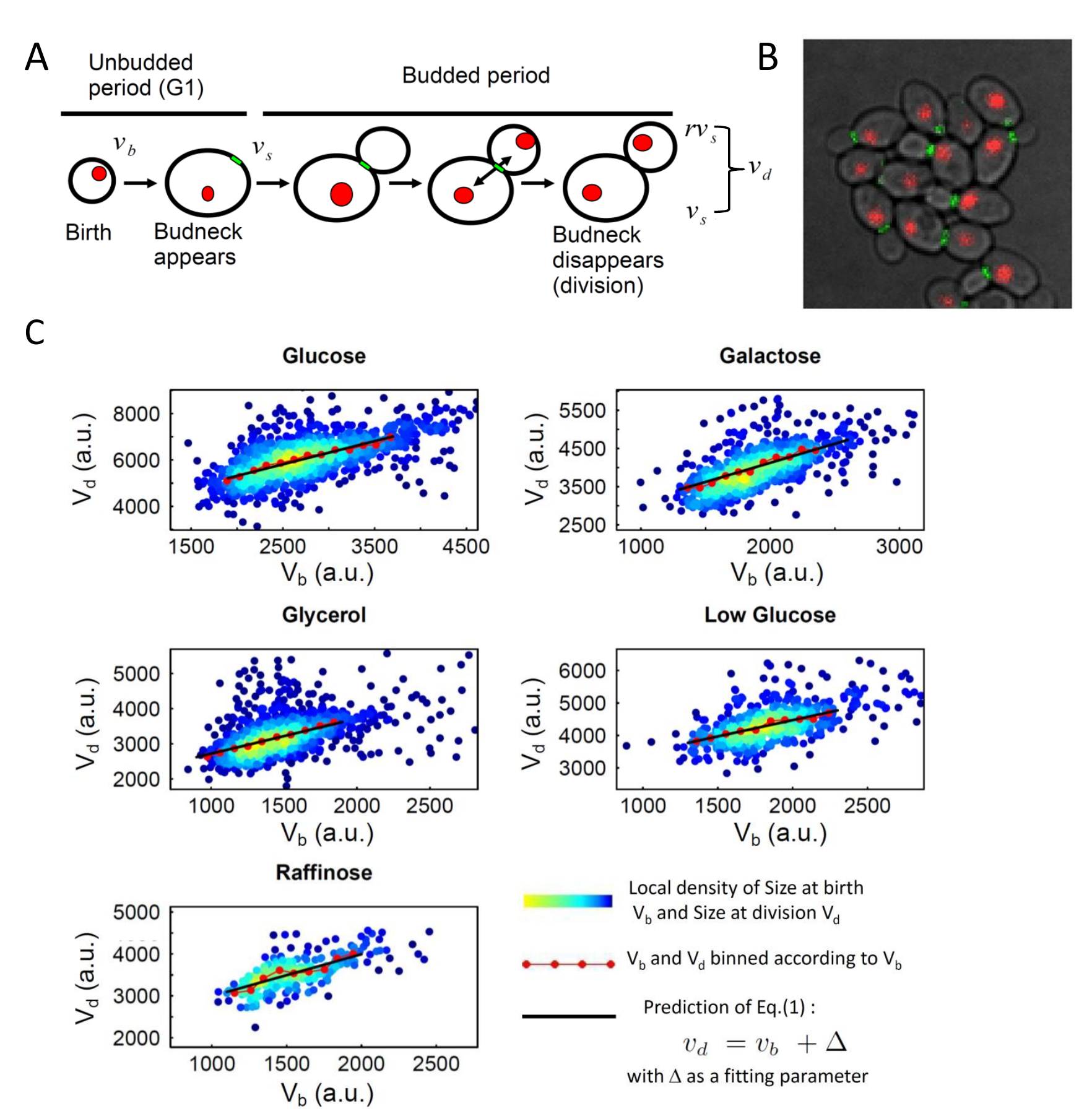}

 \caption{A \textbf{Illustration of the budding yeast cell cycle} Fluorescent labeling of bud neck ring and the nucleus by the fusion proteins Cdc10-GFP \cite{bean} and Acs2-mCherry \cite{huh}, respectively, enabled precise definition of cell birth, cell division and the initiation of budding (see SI section 1).The diagram shows the Start transition from G1 to the budded phase, and defines the notation used in the text for the different cell cycle variables.
B \textbf{Growing microcolonies of \emph{S. cerevisiae}} A typical image from our experiments on budding yeast: overlay of a brightfield picture and green and red fluorescent images. The strain used carries a fluorescent marker of the bud neck ring (Green; indicates cell division), and a marker of the nucleus (Red; assists the identification of cells in the image processing algorithm).
 C \textbf{Positive correlations between size at birth and division in \emph{S. cerevisiae} daughter cells}. Size at birth and division of daughter cells, grown in 5 different conditions. The color of the dots (blue to yellow) represents the local density. Red dots: data binned according to the size at birth. The black lines show the predictions of Eq. (\ref{incremental}) -- for all growth conditions the slope of the plotted line is 1, and the offset $\Delta$ is taken as a fitting parameter (different for each growth condition, since the average cell size depends on the growth medium). A linear regression analysis on the raw data for the 5 different growth media yields a slope of 0.91, 0.99, 0.95, 1.01 and 1.1 for glucose, galactose, glycerol, low glucose and raffinose respectively, showing excellent agreement with prediction A.}
\label{size_size}
\end{figure}

It is appealing to interpret Eqs. (1-2) as indicating that the cell measures and controls the volume added between birth and division. However, our data makes this model unlikely. As explained in the SI (section 2 and Fig. S3), our data shows that the duration of the budded phase is uncorrelated with the cell size at birth. If the cue for division is the accumulation of a constant volume from birth, it is not clear how the cell could initiate budding such that division would occur a constant time later: this would be akin to measure a constant \emph{negative} time from division. Moreover, previous research indicates that size control in budding yeast occurs at the G1/S transition (roughly speaking, the onset of budding) \cite{review2}, rather than at division. We found that a solution to this seeming paradox can be achieved if we avoid the interpretation of Fig. \ref{size_size} as control of the volume accumulated between birth and division, and instead consider a model where the control is over \emph{budding}. Our model makes the following assumptions:

1) Division occurs a constant delay after Start. This assumption is in agreement with the lack of correlations between the budded phase duration and size which we observed in our data (Fig. S3).

2) During the budded phase of yeast, almost all of the cell growth occurs in the bud -- which describes the experimental observations well (the ratio of the mean cell size at budding to mean parent cell size at division is between 1.017-1.05 in our experiments).

3) A constant volume increment is added between two budding events. In the SI (sections 3 and 5) we show that if budding is triggered by accumulation of sufficient copies of an initiator protein, produced in proportion to volume growth and partitioned between the two cell bodies in relative proportion to their volume, then Eqs. (1) and (2) follow, as well as predictions C-D discussed below. However, we found that the model is also mathematically equivalent to an inhibitor model (SI section 4), in which budding occurs when the level of an inhibitor falls below a critical level: if a constant number of inhibitor molecules are produced in G2, and are partitioned between bud and parent cell in relative proportion to their volume, all correlations are identical to those of the initiator model (i.e., both models lead to a constant volume increment between two budding events). Interestingly, this model is similar to the model proposed in Ref. \cite{skotheim}, where the dilution of the protein Whi5 as the cell grows is suggested to be responsible for size control. This appears to be a plausible molecular mechanism to implement the incremental model in budding yeast.

In section 5 of the SI we show that assumptions 1-3 (in either interpretation, of an initiator/inhibitor model) lead to the following predictions:

(A) Adding a constant volume between two Start events leads, non-trivially, to Eq. (1): hence plotting size at division versus size at birth is expected to produce a linear relationship with slope 1.

(B) Eq. (\ref{incremental}) implies a negative correlation between interdivision time and size at birth, specifically, Eq. (\ref{time_corr}).

(C) The assumption of a budded phase of constant duration implies a division asymmetry (bud:parent cell ratio) independent of size at birth; thus there would be no correlations when plotting the asymmetry against the size at birth.

(D) We find that the volume increment during budding is \emph{positively} correlated with the size at birth, with a slope of $\frac{r}{1+r}$, while that during $G_1$ is \emph{negatively} correlated with it with a slope of $-\frac{r}{1+r}$, where $r$ is the bud:parent cell volume ratio at cell division. These two contributions cancel during a full cell-cycle, leading to a constant volume added between two budding events, as shown in Fig. 3.

There is an important difference between the nature of predictions A-B as opposed to C-D: while the former cannot distinguish between adding the volume from birth to division versus adding it between two budding events, predictions C-D are inconsistent with the addition of volume between birth and division, and as such are a useful way to distinguish between these two cases. It should also be emphasized that these predictions are different from those associated with a critical size model: for example, in that case there would be no correlation between cell size at birth and division. Our results suggest that control acts at the G1/S transition, in agreement with other studies \cite{review2,siggia,hartwell,lord}, yet the particular mechanism we suggest is different than the current paradigm.

Figs. 1-2 show the excellent agreement of predictions A and B with our data. The agreement of prediction C with the data is shown in the SI, where we found the division asymmetry $r$ to be uncorrelated with the cell size at birth (Fig. S3). The comparison of prediction D with our data is shown in Fig. 3, showing the \emph{lack} of correlations between volume added between two consequent budding events and size, and in Fig. S3, showing the positive and negative correlations between volume increment and size when considering the G1 and the budded phase, respectively. The agreement of the experimental results with the model was good in 4 different culture media. It was  poorer when cells grow on raffinose as a carbon source, for reasons which we do not understand.

\subsection*{The incremental model in bacteria}

In order to emphasize the striking similarity of the size control strategy in yeasts and bacteria, we performed the same analysis on data previously collected with \emph{E. coli} by Stewart et al. \cite{stewart} and Wang, Robert et al. \cite{jun} (details regarding the data sets and the analysis can be found in SI, section 7). Fig. 4A shows that the same correlations as we showed for budding yeast in Figs. 1-2 (corresponding to predictions A and B) also describes the data in \emph{E. coli}. Section 3 of the SI discusses a potential molecular mechanism which appears to be relevant for bacteria, where an initiator protein is accumulated between two DNA replication initiation events -- akin to the budding-to-budding volume accumulation in budding yeast. In addition to reproducing the experimentally observed correlations, this model explains the known exponential dependence of size on growth-rate, as shown in Ref. \cite{amir_PRL_size}, and, importantly, regulates the number of multiple replication forks in addition to controlling size \cite{ho}.

\subsection*{Size and time distributions can be collapsed for bacteria and yeast}

An additional prediction of our model is that a \emph{single} parameter describing the noise magnitude will determine both the size and interdivision time distributions, as derived in the SI (section 6). This allows us to scale the experimentally measured distributions for size and interdivision time. Fig 4B (left) compares the theory with the experimental results for budding yeast growing in glucose, showing the excellent scaling collapse obtained in this way.
An important property of these distributions is their coefficient of variation (CV). A-priori, one may think that the CV of the size distribution is independent of that of the interdivision time distribution. However, since a single source of stochasticity is responsible for the widths of both distributions in our model, the ratio of the CVs is uniquely determined.
For \emph{E. coli}, the assumption of nearly symmetric division simplifies the calculations and allows to obtain analytic formulas which are not possible for asymmetric division.
 As predicted in Ref. \cite{amir_PRL_size}, the distribution of size at birth is relatively narrow and its CV is $\log(2)\approx 0.69$ smaller than that of the interdivision times. In agreement with this prediction, we estimated the ratio of CVs of size at division and interdivision time in the data set from Wang, Robert et al. and found $0.69 \pm 0.03$ for 3 independent experiments. The theory thus predicts that the distribution of the normalized logarithm of size at birth, $\log_2(v_{b}/v_{0})$ (with $v_0$ the average size at birth), should collapse on the distribution of interdivision time appropriately rescaled, $(t/\tau_d-1)$. Fig. 4B (right) shows the distributions of size at birth and interdivision time normalized according to the theory, the excellent collapse of the curves supporting the validity of our stochastic model.

\section*{Discussion}

In this work, we showed that size control in budding yeast relies on an incremental strategy, leading (effectively) to the addition of a constant volume between birth and division. This is in contrast to the long standing paradigm where the Start transition occurs when the cell size reaches a threshold value. Our study of correlations shows that the incremental strategy is likely to be implemented at the G1/S transition: the cell adds a constant volume between two Start transitions, not between birth and division.

Despite the differences in morphology, DNA replication and growth of \emph{S. cerevisiae} and \emph{E. coli}, we showed these two organisms control their size using an identical strategy -- described mathematically by the incremental model, where a constant volume is added between two events in the cell cycle. The correlations between size at birth and at division, and between size at birth and interdivision time are quantitatively predicted by this model, and agree well with our experimental data, for both organisms.
In bacteria, DnaA is known to be a key regulator of the cell cycle, as it triggers initiation of new rounds of DNA replication. Similarly, in budding yeast Whi5 seems to be a leading candidate in cell cycle regulation. Thus, similar cell cycle control in both organisms is likely a result of convergent evolution rather than of an identical molecular mechanism. In fact, even the principles by which the molecular mechanisms operate may be different in the two; In the implementation of the incremental model first introduced by Sompayrac et al. to describe the bacterial cell cycle \cite{incrementalmodel}, a size increment is added between two successive events of DNA replication initiation, through the accumulation of an initiator. Division then occurs after a constant delay, leading to a constant increment of volume between birth and division. On the other hand, the molecular mechanism in \emph{S. cerevisiae} may be due to the dilution of the inhibitor Whi5, a model supported by recent experiments \cite{skotheim}. We have shown here that this model implements the incremental model, if the inhibitor is shared between the parent cell and the bud in proportion to their volume. The simple mechanism which we propose explains both the correlations that we observed, and is in accord with the widespread view that in budding yeast size control occurs via control at the G1/S transition \cite{johnston1977coordination,review2}.

An appealing feature of our model is that it offers a coordination of different events in the cell cycle, namely growth, division and DNA replication. DNA replication is coupled to growth, size control acting at the level of initiation of replication/Start, and is also coupled to division. Our work paves the way for an improved molecular level understanding of size control, and combining our phenomenological observations with the molecular techniques as used in Refs. \cite{siggia, tang, skotheim} is a promising direction. It would also be interesting to repeat the analysis we applied here for \emph{E. coli} and \emph{S. cerevisiae} to other organisms and find the regime of applicability of the incremental model, which will shed new light on the cell cycle, for both eukaryotes and prokaryotes.

\section*{Experimental procedures}

  We used high throughput time lapse microscopy with a built-in auto-focusing apparatus \cite{paran} and developed automatic software that enabled tracking individual cells over multiple division cycles. Our analysis identified timings of cell cycle transitions and respective cell volumes. We grew yeast cells at different growth rates by changing the carbon source in the medium (glucose at high or low concentrations, galactose, glycerol and raffinose). Yeast strains were as described in Ref. \cite{soifer2014systematic}.

\subsection*{Yeast time-lapse microscopy}

Yeast cells were pre-grown for around 24 hours in SC medium to OD600 of about 0.5. The carbon sources used were as follows: 2\% glucose, 2\% galactose, 0.05\% glucose, 2\% raffinose and 2\% glycerol + 2\% ethanol. The cells were then prepared for imaging growing on agar pads with the respective SC as previously described.  We observed growth of microcolonies at 30C using fully automated Olympus IX71 inverted microscope equipped with a motorized XY and Z stage, external excitation and emission filter wheels (Prior) and an IR-based fast laser autofocus \cite{paran}.  Fluorescent proteins were detected using EXFO X-Cite light source at 12.5\% intensity and Chroma 89021 mCherry/GFP ET filter set. Exposure time for the detection of eGFP and mCherry was 120 msec. Imaging was done by cooled EMCCD camera (Andor). The microscopic setup allowed simultaneous imaging of 60 fields of view for 6 hours. Bright field, red and green fluorescence images were collected every 3 minutes for the fermentable and every 5 minutes in the non-fermentable carbon sources.

\subsection*{\emph{E. coli} data analysis}
We analyzed the results of video-microscopy experiments performed by Stewart et al. \cite{stewart} and Wang, Robert et al. \cite{jun}. See SI section 7 for the details of data analysis.

\section*{Acknowledgements}
We are indebted to Naama Barkai and members of her lab for support during
development of the project. We thank Andrew W. Murray for stimulating discussions
and to Eric Stewart for sharing his data. A.A. acknowledges the support
of the Harvard Society of Fellows, the Milton award, and the Sloan Foundation.

\newpage

\pagestyle{empty}

\begin{figure}[h]
\includegraphics[scale=0.75]{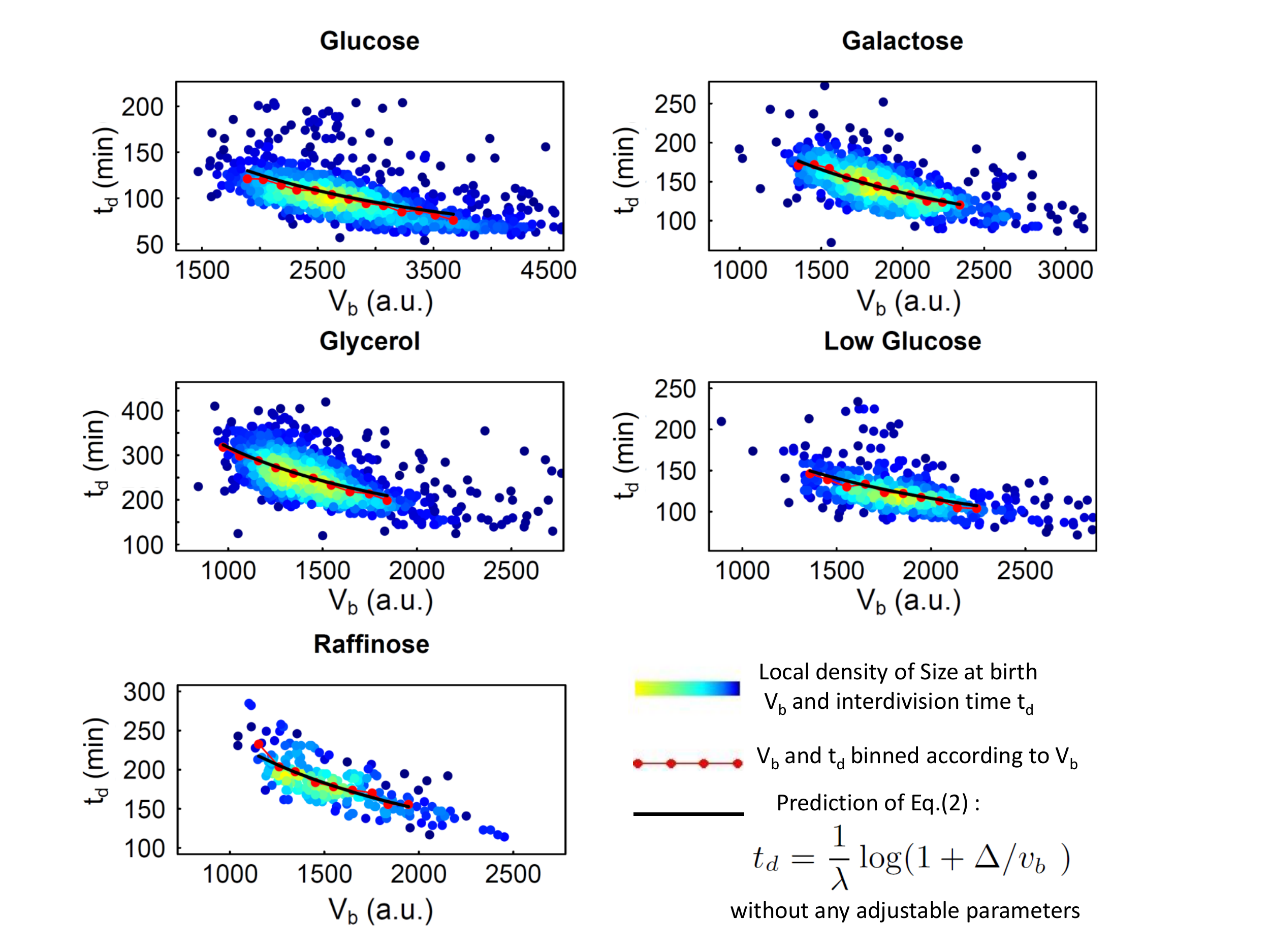}

 \caption{\textbf{Negative correlations between size at birth and interdivision time in \emph{S. cerevisiae} daughter cells.}
Size at birth and interdivision time of daughter cells, grown in 5 different growth conditions. The color of the dots (blue to yellow) represents the local density. Red dots: data binned according to the size at birth. The black lines show the theoretical prediction B (Eq. (2)), without any adjustable parameters. }

\label{size_time}
\end{figure}

\begin{figure}[h]
\includegraphics[scale=0.75]{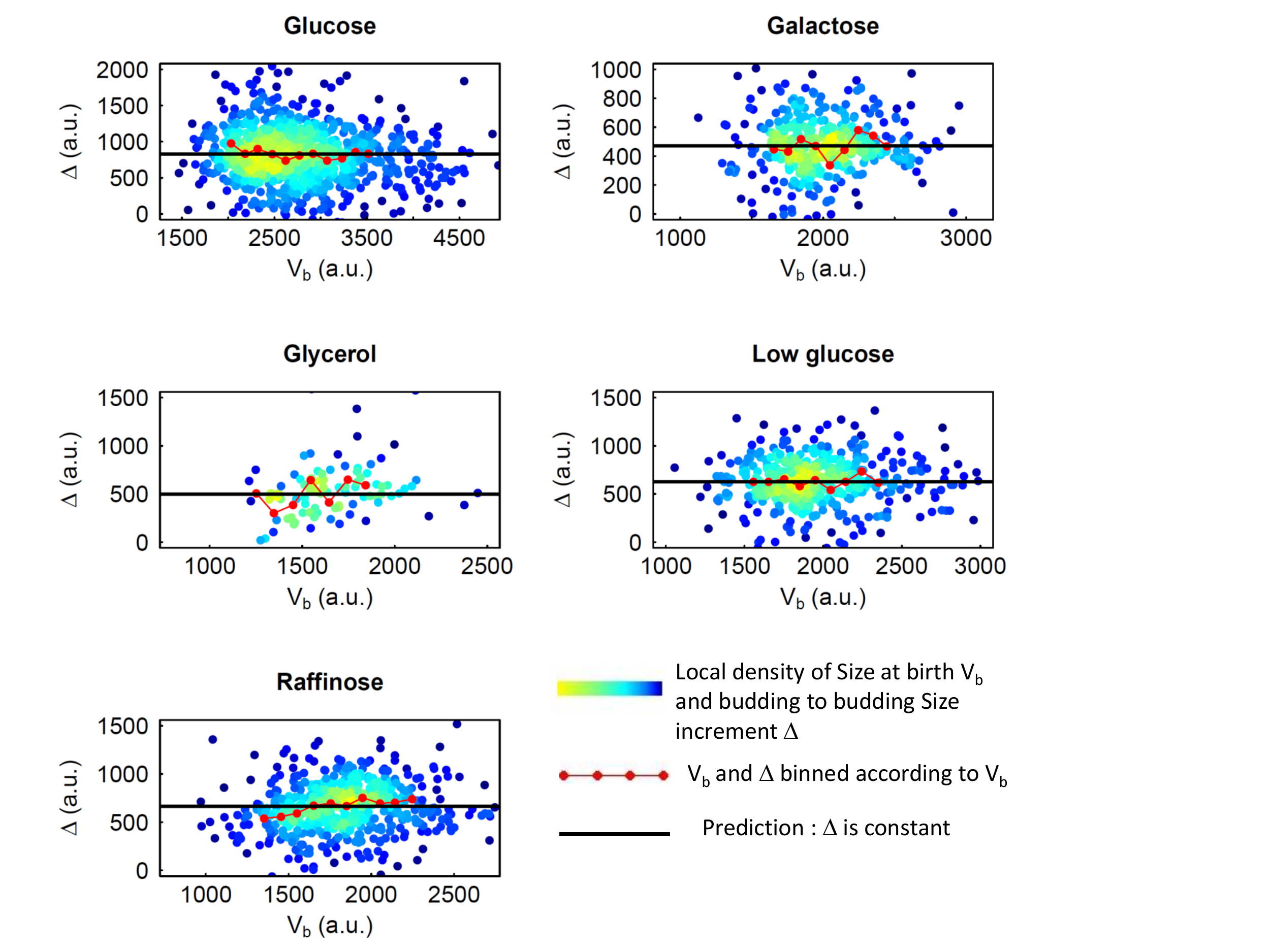}

 \caption{\textbf{No correlations between budding to budding volume increment and size at birth.} We found no correlation between the total volume added between two budding events and the size at birth, as expected from our model: the positive correlations of the increment during the budded phases (Fig. S3B) and the negative ones during G1 (Fig. S3A), cancel out to give the incremental model between two budding events. Note that in light of the mechanisms proposed in sections 3 and 4 of the SI, the volume increment during the budded phase is assumed to divide between the bud and parent cell in proportion to their relative volume. }
\label{budding_budding}
\end{figure}

\begin{figure}[ht]
\vspace{-1 cm} \includegraphics[scale=0.7]{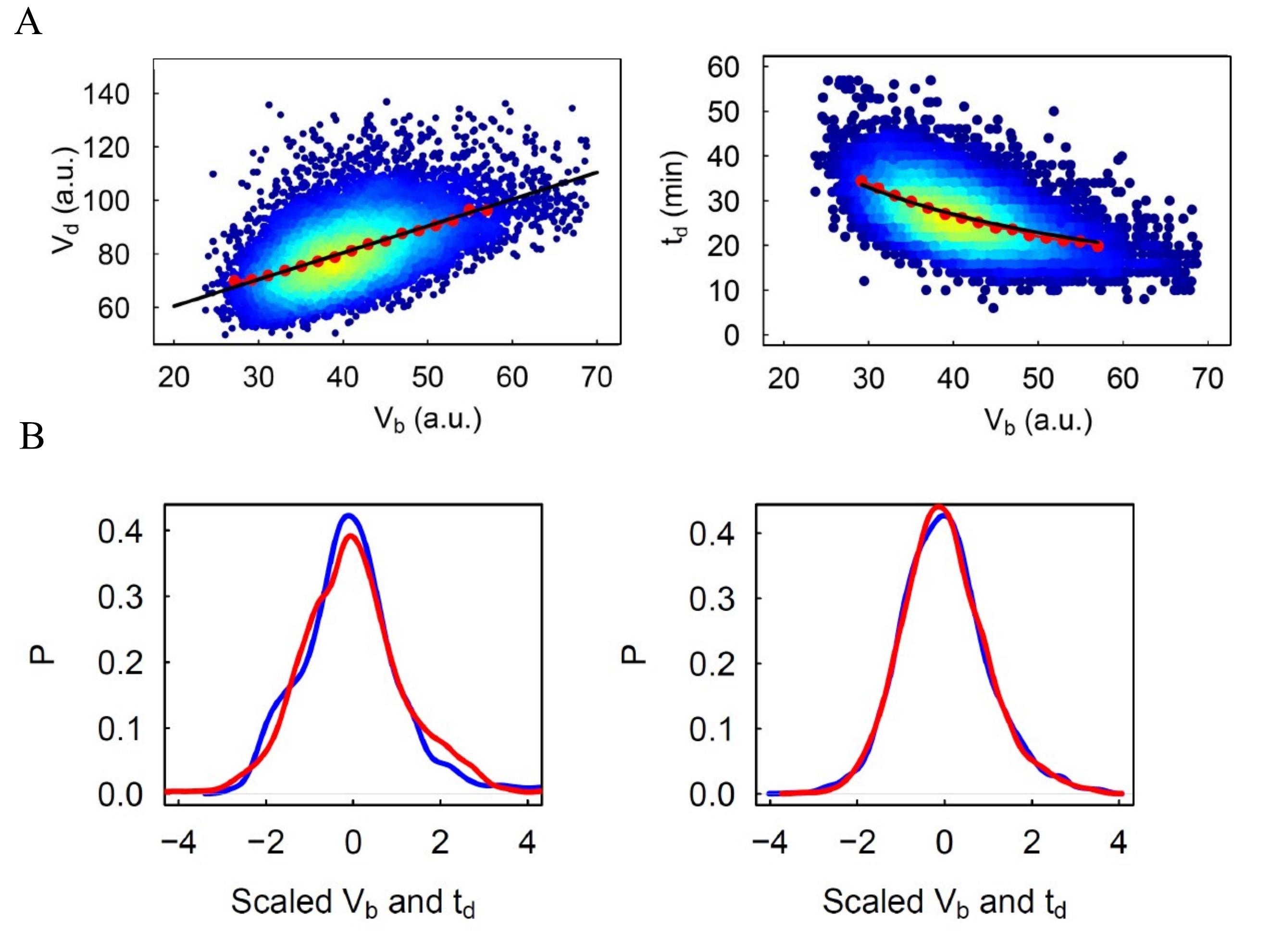}

\caption{A \textbf{Correlations between size at birth, size at division and interdivision time support incremental model in \emph{E. coli}} (left) Size at birth and division of single cells in fast growth conditions (Stewart et al. data set; see SI for more information). The color of the dots (blue to yellow) represents the local density. Red dots: data binned according to the size at birth. The black lines show the predictions of Eq. (\ref{incremental}), with a slope 1. In this case, the offset $\Delta$ is equal to the average cell size at birth, since the division is approximately symmetric.  (right) Size at birth and interdivision time of single cells in fast growth conditions. The favorable comparison of the binned data (red points) with the prediction of Eq. (\ref{time_corr}) (solid line), with no fitting parameters, strongly supports the incremental model for size control in \emph{E. coli}.
B \textbf{ Distributions of size at birth and interdivision time can be scaled.} (left) Distribution of newborn size of yeast daughter cells in glucose, and the interdivision time distribution, are scaled according to the theory (see SI for details). The only fitting parameter in the theory is the magnitude of a stochastic noise, $\sigma_T$, which accounts for the coefficient of variation of \emph{both} distributions. (right) Similarly, for symmetric divisions, the incremental model predicts that the size distribution is narrower than that of the interdivision time distribution by $\log(2)$. This implies that the distribution of $\log_2(v_{b}/v_0)$ (with $v_0$ the average size at birth), should collapse when plotted against the distribution of  $(t-\tau_d)/\tau_d$ (and $\tau_d$ the doubling time). This is shown in the figure, where the blue line is the scaled size distribution $\frac{1}{C}\log_2(v_{b}/v_0)$, and the red line is the scaled time distribution $\frac{1}{C}(t-\tau_d)/\tau_d$, with $C =0.2$  (note that the collapse is independent of the choice of $C$). All distributions were generated from the data using a kernel density estimation. }
\label{ecoli}
\end{figure}

\newpage
\renewcommand{\thefigure}{S\arabic{figure}}

\setcounter{figure}{0}
\huge{\textbf{Supplementary Information}}

\normalsize

\section{Analysis of budding yeast data}
\begin{figure}[h!]
\center{\includegraphics[width=1\linewidth]{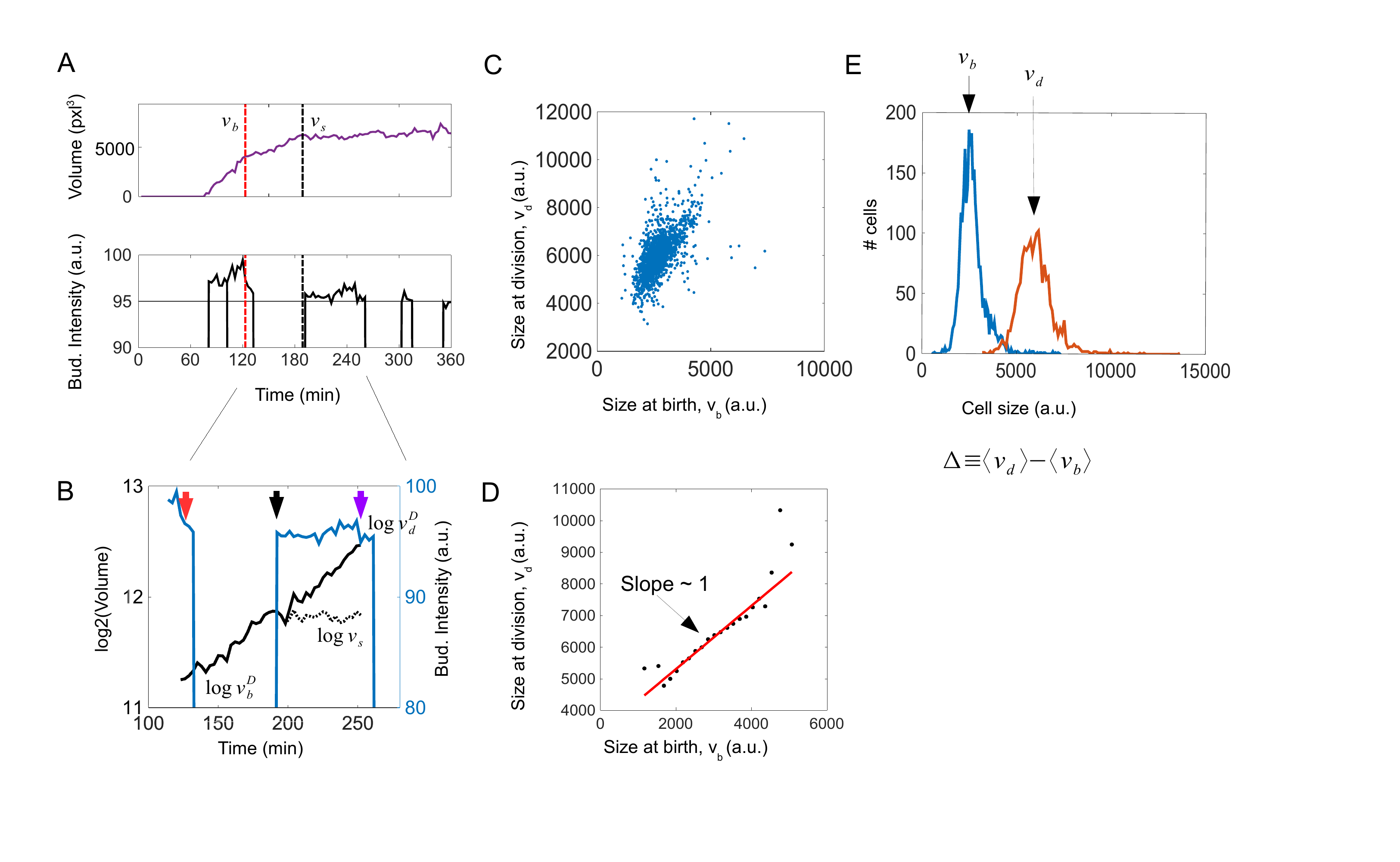}}

\caption{\textbf{Illustration of the budding yeast data analysis, related to Figure 1}.  (A-B) Image analysis identified events at cell cycle. A Top panel: volume growth of one cell body, starting from appearance at time 70, detachment from parent cell (red dashed line at time 120'), growth during G1 (until bud neck appearance at time 190', black dashed line) and constant volume of the cell body after bud emergence. Bottom panel shows the corresponding intensity of the bud neck on this cell body.
(B) Parent cell and bud relationships were identified using bud neck marker. Combined volume grows exponentially (solid black line), while parent cell body remains of constant volume (dashed line). Red, black and violet arrows mark  parent cell birth, bud emergence and detachment of the first daughter cell respectively. Total volume of the parent and bud cell bodies is denoted $v_d$. (C-E) Statistical procedure for estimation of slopes and $\Delta$. (C) raw volumes at birth and division show significant correlation. (D) Raw data was binned into equally spaced bins of birth sizes, linear fit was fitted to the binned line and showed slope close to 1. (E) The constant volume increment between birth and division is denoted $\Delta$ and was calculated as the difference between mean size at division and at birth. }
\label{yeast_analysis}
\end{figure}

Yeast strains were grown on agar pads containing synthetic complete medium placed on a coverglass. This allows for unperturbed growth in a planar fashion for 3-4 generations. Our microscopic setup enabled collection of up to sixty fields of view every three minutes for six hours. Details of the microscopic setup and the image analysis procedure were previously described \cite{soifer2014systematic}. Fluorescent markers enabled identification of three points in the cell cycle (Fig. \ref{yeast_analysis}A): cell birth, marked by disappearance of the bud neck, beginning of S phase (marked by appearance of the bud neck) and metaphase to anaphase transition: splitting of the nuclear marker. Here we are predominantly interested in the former two events, using the latter as a facilitator of image analysis.
Image analysis software (Fig. S\ref{yeast_analysis}A) segmented and tracked cell bodies (i.e. parent cell parts and buds separately) starting from emergence of the cell body as a bud. Nuclear separation marked appearance of the nucleus in the cell body. Nuclear separation was followed by disappearance of the bud neck (dashed red line on Fig. S\ref{yeast_analysis}A) marking cell birth. Volume at this point (i.e., volume of the newborn daughter cell) was denoted by $v_b$. The daughter cell continued to grow during G1, at the end of G1 (appearance of the bud neck, dashed black line on Fig. S\ref{yeast_analysis}A) the growth of the cell body stopped as all the cell growth occurs in the bud. Size of the daughter cell at bud emergence is denoted $v_s$.

Bud neck connection allowed assignment of parent cell to each bud. this allowed calculation of the total size of the cell at division: size of the parent cell body plus the size of the  bud (Fig. S\ref{yeast_analysis}B, note log scale), this size was denoted $v_d$. The average difference $v_d-v_b\equiv\Delta$.

$v_d$ and $v_b$ were strongly positively correlated (Fig. S1C). To calculate the slope of the correlation without a significant effect of the outliers, we binned the $v_b$ into equally spaced bins and calculated mean $v_d$ of each bin. The slope of this line was close to one in all cases examined for the diploid strain (Fig. S\ref{yeast_analysis}D). Finally, $\Delta$ was calculated as the difference between means of sizes at division and at birth (Fig. S\ref{yeast_analysis}E).

\newpage

\newpage

\section{Exponential growth at the single-cell level and a constant asymmetry ratio imply a constant duration of the budded phase}

Exponential growth at the single-cell level implies a linear dependence between $\log(v_d/v_b)$ and time, which we verify in Fig. S2A. Recovering this known result \cite{siggia} is evidence to the accuracy of the measurements and the analysis.

Yeast cell asymmetry (daughter cell volume divided by parent cell volume) is \emph{uncorrelated} with the cell size (Fig. S2B). Using the exponential nature of the growth at the single-cell level, the ratio of daughter to parent cell size  just after division is given by:
\be r = e^{\lambda t_b} -1 , \label {asym} \ee where $t_b$ is the duration of the budded phase and $\lambda$ is the growth rate (e.g.: symmetric division would correspond to $r=1$). The fact that $r$ is uncorrelated with cell size implies that $t_b$ is independent of the size at birth or at budding.

\begin{figure}[h!]
A   \hspace{10 cm} B \\
\includegraphics[width=0.5\linewidth]{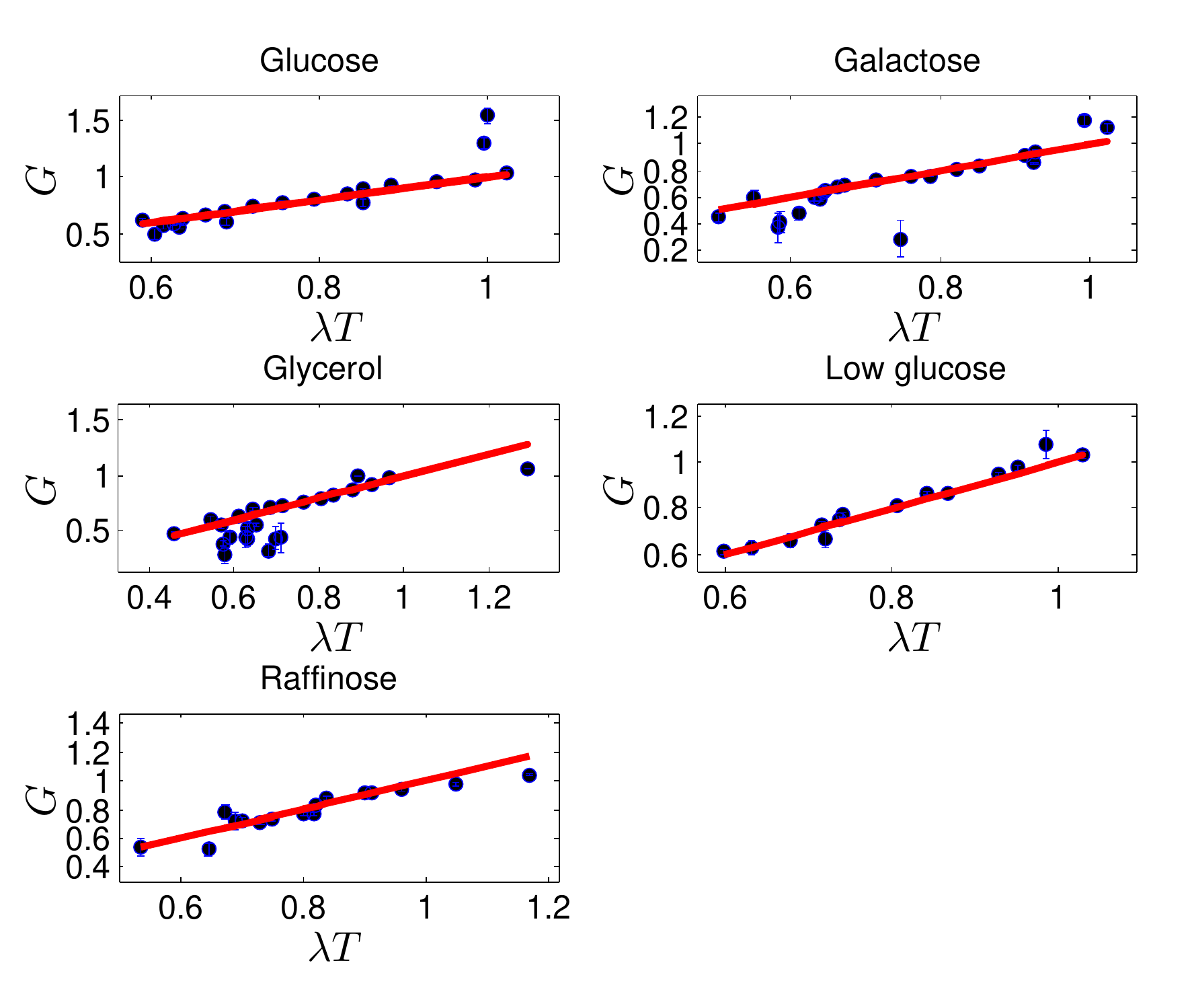}
\includegraphics[width=0.5\linewidth]{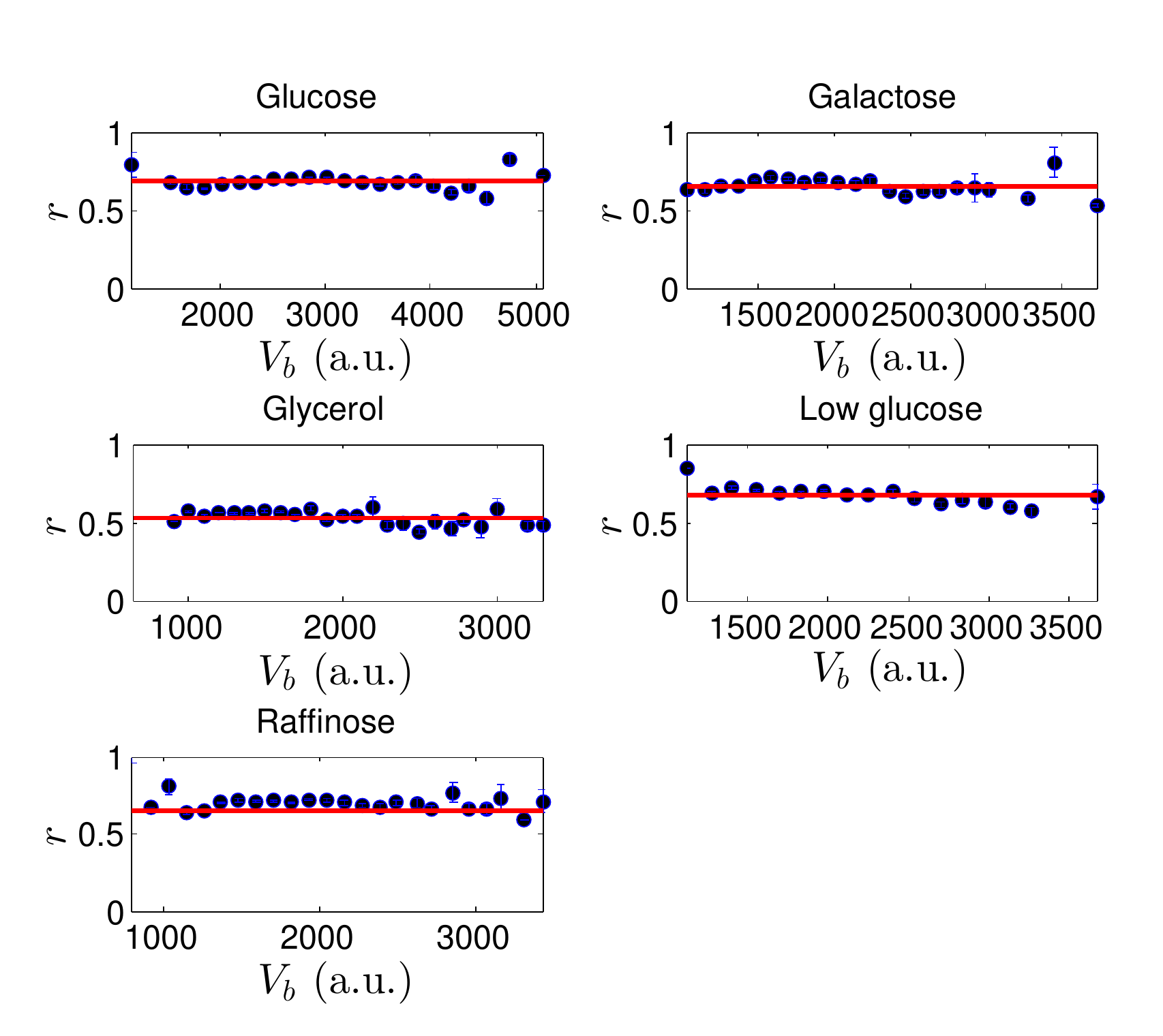}

\caption{\textbf{A Exponential growth of cell volume, related to Figure 1}. The relative growth between daughter cell birth and division, $G \equiv \log(v_d/v_b)$, is shown to be equal to the interdivision time $T$ normalized by the growth rate, $\lambda$. This linearity, shown on binned data in 5 different growth media, is consistent with exponential growth at the single cell level and would be in sharp contrast to a linear growth law. \textbf{B Budding yeast have a constant asymmetry ratio.}  The asymmetry of division $r$, related to the time $t_b$ of the budding phase via Eq. (\ref{asym}), is uncorrelated with the cell size at birth. The ratio of daughter to parent cell size at birth $r$ is estimated on 1000-2000 dividing daughter cells and binned according to their size at birth (blue dots; bins with less than 4 points are ignored). The error bars are standard error of the means. See the Experimental Procedures section in the main text and SI section 1 for further details regarding the experimental setup.}
\label{test_assumptions}
\end{figure}

\section{Implementation of the incremental model through accumulation of an initiator}

A hypothetical model that may implement the incremental model in bacteria has been proposed already in 1973, in Ref. \cite{incrementalmodel}. For completeness, we reproduce the key arguments:
One protein autoregulates, such that it operates at a constant concentration (i.e., when its concentration increases, it will autorepress more strongly and lower it, and vice-versa, leading to a stable concentration approximately independent of size or growth-rate). Another protein, the ``initiator", is found in the same operon such that it is produced concurrently with the autorepressor, and localizes at the origin(s) of replication. When the number of initiators at the origin  reaches a threshold value, DNA replication will be initiated, upon which the initiator will be degraded.

Let us analyze the behavior of this control mechanism: upon a volume increase $\Delta V$, the amount of autorepressor produced will be proportional to $\Delta V$, to maintain its constant concentration. The same is thus also true of the initiator, but since its number is ``reset" at initiation, this implies that the \emph{copy number} of the initiator is proportional to $\Delta v$. Therefore, by thresholding the copy number of the initiator per origin we will be thresholding $\Delta V$ per origin -- therefore implementing the incremental model.

This is, first and foremost, a proof-of-principle of a simple molecular mechanism which may implement the incremental model, and elucidates what we mean by ``volume integration". Within this model it is clear that upon cell division the ``volume increment" (i.e., the amount of the initiator) should be divided equally between the two daughter cells (for symmetric division). Interestingly, the protein DnaA shares much in common with this hypothetical mechanism: it is known to autorepress \cite{wright}, and has two forms -- an ADP-bound and ATP-bound form. The latter form localizes at OriC, and 20 copies of it are needed to initiate DNA replication, upon which it is converted to the ADP-bound form \cite{levin}.

Ref. \cite{amir_PRL_size} shows that this model leads to an effective correlation between birth and division identical to the incremental model:

\be v_d = v_b + \tilde{\Delta},\ee

but with $\tilde{\Delta}$ depending on the growth-rate as:

\be \tilde{\Delta} \propto  e^{\lambda T},\ee

with $T$ the duration of the C+D period, known to be approximately constant and equal to 60 minutes in \emph{E. coli} at 37$^o$ C.
This implies that the average cell size at birth (or division) depends exponentially on the growth-rate, which has been experimentally observed in the 1950's \cite{schaechter}, the exponent being indeed very close to 60 minutes.
Recently it was shown that the same mechanism has the important property of regulating the number of replication forks, in addition to volume \cite{ho}, suggesting that it is a simple yet highly effective mechanism to couple different events in the bacterial cell cycle. It is also consistent with the phenomenon of ``rate-maintenance" \cite{cooper_2}, namely, that the rate of initiation of DNA replication is unchanged for the first 60 minutes following a shift in the growth medium.

It is possible to repeat the calculation for asymmetric division, as is the case for budding yeast, and see that the mapping to the incremental model holds also in this case, when one assumes that the number of initiator molecules that goes to the bud/parent are divided according to the relative volume (i.e., the initiator is homogenously spread through the cell). We will show this in section 5. This provides one hypothetical model which may implement the addition of a constant volume between two budding events. However, a more plausible molecular mechanism for budding yeast relies on the dilution of an inhibitor, and is discussed in the next section.

\section{Implementation of the incremental model through dilution of an inhibitor}

In the following, we will suggest a different potential molecular mechanism, which will map precisely to the incremental model, and is very similar (albeit not identical) to a recently proposed molecular mechanism for size control in budding yeast \cite{skotheim}, where dilution of the cell cycle inhibitor Whi5 controls the budding-yeast cell size. Our approach will suggest that this mechanism could be consistent with the observed correlations, but only when the details of the model are appropriately modified, as we shall now explain.

Within the model, a constant number of Whi5 molecules is produced during the budded phase. These are never degraded, but rather are diluted in G1 as the cell grows. Whi5 inhibits Cln3 (whose concentration is assumed to be constant and independent of size \cite{skotheim}), and budding will occur when its concentration is sufficiently low (i.e., after the cell has sufficiently grown to dilute the Whi5). This implies that if a cell is born with $x$ Whi5 molecules, the volume at budding will be proportional to $x$ (up to fluctuations arising from the noise). Finally, we will denote the fraction of Whi5 molecules that go to the daughter cell by $\alpha$. The simplest expectation would be $\alpha =\frac{r}{1+r}$, with $r$ is the ratio between the size of the bud to the parent cell, which corresponds to the case where Whi5 molecules are distributed homogenously throughout the cell. Summarizing, the assumptions of the model are:

1) Volume at budding is proportional to the number of Whi5 molecules. We will choose the proportionality constant to be 1, for convenience.

2) A constant number of Whi5 molecules, $V_0$, is produced in the budded phase.

3) A fraction $\alpha=\frac{r}{1+r}$ of Whi5 molecules goes to the bud at cell division.

Denoting the number of Whi5 molecules in the daughter cell birth at the i'th generation by $X_i$, we therefore have:

\be x_{i+1} = (x_i + V_0)\alpha . \label{inhibitor_eq} \ee

The volume at budding during the i'th cell cycle would be $x_i$, by assumption (1). Thus the volume at division would be $V^i_d = x_i (1+r)$,
and the cell size at birth in the next division would be:

\be V^{i+1}_b = x_i r. \ee

Therefore we find that:

\be V^i_d  = (x_{i-1}+V_0) \alpha (1+r) = V^i_b \frac{\alpha (1+r)}{r} + V_0 \alpha (1+r). \ee

In our model, Whi5 molecules are uniformly distributed throughout the cells and $\alpha =\frac{r}{1+r}$ (division of the Whi5 molecules is proportional to the relative volume). This would precisely implement the incremental model:
\be V^i_d = V^i_b + V_0 r. \label{dil}\ee

 Interestingly, Ref. \cite{skotheim} claims that more Whi5 goes to the daughter cell than the proportional volume, which would suggest a larger value of $\alpha$, corresponding to weaker size control in daughter cells -- contradictory to our observed correlations.

\section{Correlations between cell-cycle variables}
We denote the daughter size at birth $v_{b}^{D}$, see Fig. 1A of the main text, which illustrates the cell cycle in budding yeast, and defines the notations which we will use in the following calculations. Before the bud detaches, i.e. the cell divides, the size of the entire cell (parent cell + bud) is denoted by $v_{d}^{D}$ for the daughter cell. We emphasize that after budding occurs, practically all volume growth will occur in the bud and not the parent cell. Consider a cell where budding happened to occur at volume $v_{s}$. According to our assumption, after division the daughter cell size will be $v_{d}^{D}=r v_s$. We will now find the volume at the next G1/S transition, and show that the result is the same both for the model discussed in section 3 where an initiator protein accumulates as the cell grows, and in the dilution model presented in section 4.

\textbf{Initiator model}

Consider first the initiator model. The volume that has accumulated during the budded phase is proportional to $r v_{s}$, and hence $\frac{r^{2}}{1+r}v_{s}$ will go to the daughter, and $\frac{r}{1+r}v_{s}$ go to the mother. If we denote the total amount of volume needed by $\tilde{\Delta}$, the volume at which Start will happen for the daughter cell, $v_{s}^{new}$, will obey:

\be (v^{new}_s - r v_s)+ \frac{r^2}{1+r} v_s = \tilde{\Delta}. \ee

Hence:

\be v^{new}_s = \tilde{\Delta} + v_s \frac{r}{1+r}. \label{vstart} \ee

\textbf{Inhibitor model}

In this case, as explained in section 4 the volume at Start is proportional to the number of inhibitor molecules. Assuming that the inhibitor divides between mother and bud in proportion to their relative volume, we have $\alpha=\frac{r}{1+r}$, and Eq. (\ref{inhibitor_eq}) implies that:

\be x_{i+1} = (x_i + V_0)\frac{r}{1+r} .\ee

Since $v_s \propto x$, this relation leads immediately to Eq. (\ref{vstart}), also for the inhibitor model, with $\tilde{\Delta} = V_0 \alpha$. The equations which we shall now derive follow from Eq. (\ref{vstart}), and hence will hold in both initiator/inhibitor cases.

\textbf{Correlations between growth at G1 and the budded phase with cell size at birth}

The cell size at division will be $v_{s}^{new}(1+r)$, thus we find:

\be v^{D}_d= v^{D}_b+\Delta. \label{incremental}
\ee

This is a realization of the incremental model, with:

\be \Delta=\tilde{\Delta}e^{\lambda t_b}.\label{incremental2}\ee

 This formula is equivalent to Eq. (\ref{dil}).

The volume as Start $v_s$ is related to that at birth $v_b$ according to:

\be v_s = \tilde{\Delta} +  \frac{v_b}{1+r}. \ee
Hence the volume added during $G_1$ is:
\be \Delta^{G_1}_v = v_s-v_b = -\frac{r}{1+r} v_b + \tilde{\Delta}. \label{G1} \ee
Therefore we expect a negative correlation between the volume added during $G_1$ and the size at birth, with a slope depending on the asymmetry $r$ and an offset $\Delta$.

Similarly, the volume added during the budded phase is:
\be \Delta^{budding}_v = v_s r =v_b \frac{r}{1+r} +  r \tilde{\Delta}. \label{budding} \ee

The good agreement of these predictions with our experimental data is shown in Figs. S2A and S2B.
Note that the correlations of Eqs. (\ref{G1}) and (\ref{budding}) would not be correct if the constant volume were added between birth and division, and depend on the accumulation of the volume between two budding events, and the constant duration of the budded phase. However, these correlations cannot distinguish between a molecular mechanism which implements the incremental model by the accumulation of an initiator (section 3) or the dilution of an inhibitor (section 4).

\begin{figure}[h!]
A  \hspace{8.5 cm}   B \\
 \includegraphics[width=0.5 \linewidth]{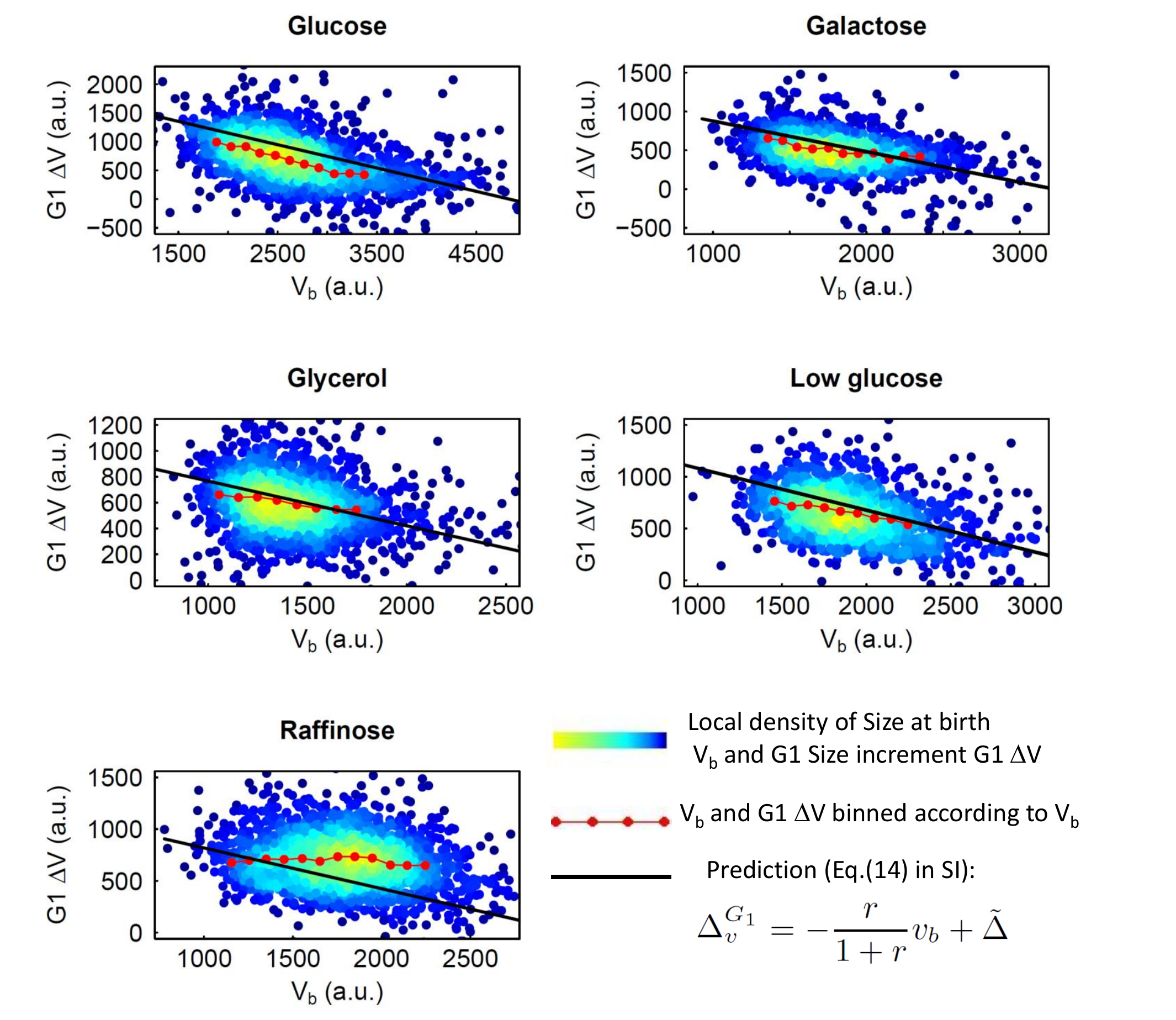}
\includegraphics[width=0.5 \linewidth]{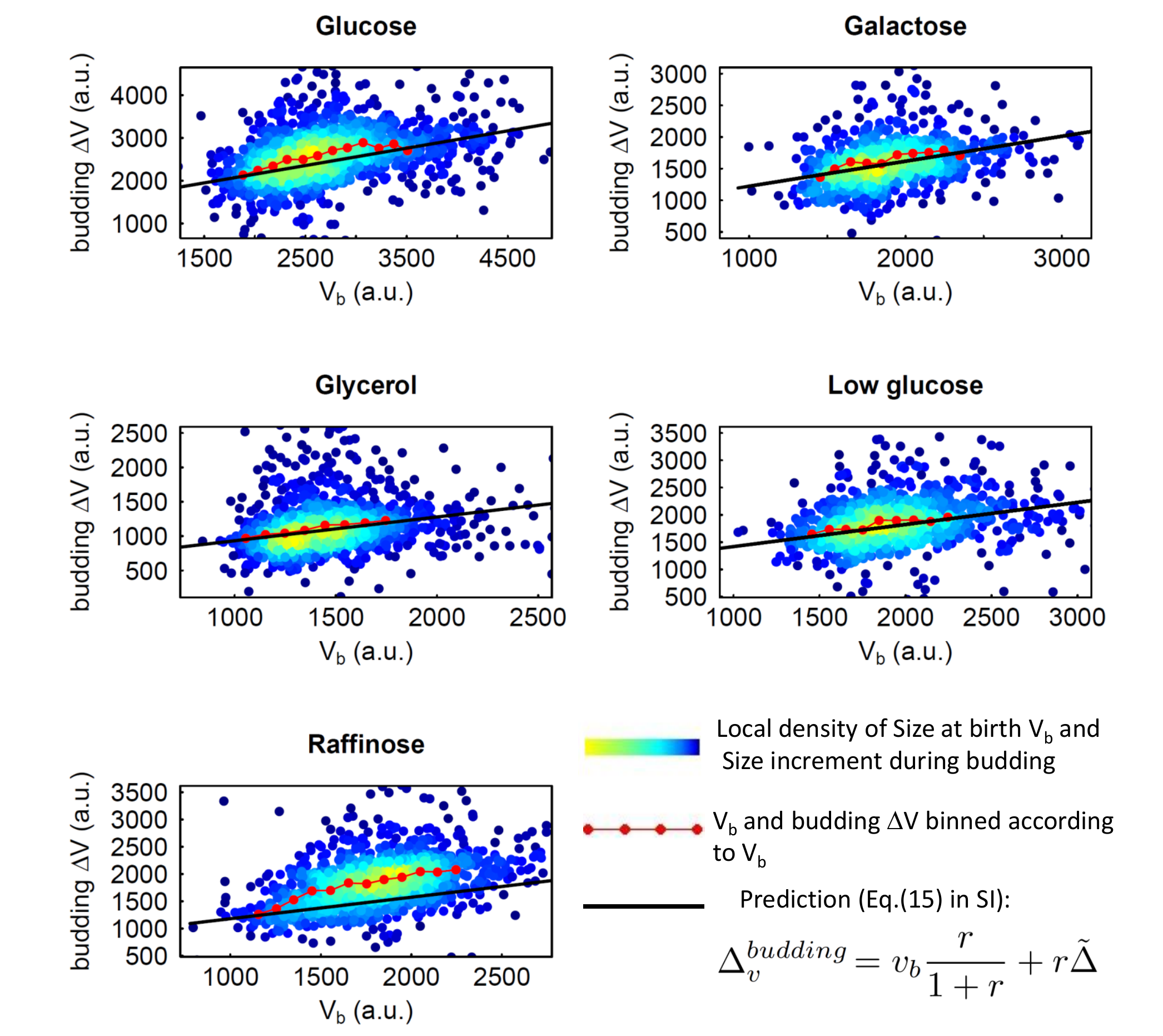}

\caption{ \textbf{Correlations between $G_1$ and budded phase durations and size, related to Figure 2}\newline
\textbf{A Negative correlations between size at birth and volume added during $G_1$ in \emph{S. cerevisiae} agree with the incremental model.} The correlation between cell size at birth and the volume added during $G_1$ is shown for diploid daughter cells grown in 5 different growth conditions. The black lines show the theoretical prediction of Eq. (14), \emph{with no fitting parameters}. \textbf{B Positive correlations between size at birth and volume added during budding in \emph{S. cerevisiae} agree with the incremental model.} The correlation between cell size at birth and the volume added during the budded phase is shown for diploid daughter cells grown in 5 different growth conditions. The black lines show the predictions of Eq. (15), \emph{with no fitting parameters}.}
\label{corr_test}
\end{figure}

In agreement with the literature, mother cells showed weaker size control \cite{review2}, manifested by correlations of cell size at birth and division with a larger slope of 1.1-1.3, as discussed in the next section. Haploid daughter cells exhibited lower slopes, of 0.7-0.8.

\section{Approximate distributions of size at birth and interdivision time for asymmetric division}

For bacteria, Ref. \cite{amir_PRL_size} shows that the distribution of $\log_2(v_b/V_0)$ should theoretically collapse with the distribution of $(t-\langle t \rangle)/\langle t \rangle$, which is related to the fact that within the incremental model the CV of the size distribution is approximately $\log(2)$ times that of the time distribution:

\be \frac{CV_{size}}{CV_{time}} \approx \log(2). \label{cv} \ee

 The right panel of Fig 4b of the main text corroborates this prediction. The purpose of this section is to find the relevant scaling within the model for the case of \emph{asymmetric} division, which will turn out to be different. The main text shows the agreement between the predictions of the incremental model and the experimental data for daughter cells (cells in the first generation after budding off), but does not discuss size control in mother cells. It is believed that daughter cells have a stronger size control \cite{review2}, which is indeed supported by our data: when considering the dependence of size at division on the size at birth, a smaller slope would correspond to a tighter size control -- for example, the ``optimal" case of a critical size mechanism corresponds to a vanishing slope. The main text shows that to a good approximation the slope is 1 for daughter cells. A linear regression for the mother cells gives 1.31, 1.21, 1.12, 1.16 and 1.26 for glucose, galactose, glycerol, low glucose and raffinose respectively. These values are sufficiently different from those of the daughter cells to support the notion that a different size-control mechanism operates in mother cells.

  We found the stationary distribution of daughter sizes numerically, implementing the incremental model for the daughter cell and taking for the mother cells a linear size control policy with a slope which we varied between 1 and 1.3. We found that changing the slope for the mother size control from 1 to 1.3 results in a change of only few percents in the ratio of the size and time coefficients of variation. For example, it changed from 0.96 to 0.93 for the model parameters corresponding to growth in glucose, which correspond to the scaling of Fig. 4b in the main text. Note that in our simulation, we assumed that the noise is \emph{added} to the interdivision time, following Refs. \cite{siggia, amir_PRL_size}. The magnitude of this noise was taken as the single fitting parameter in the model. As is shown by the data collapse of the left panel of Fig. 4b, using this single parameter we can explain the widths of both size and time distributions, supporting our stochastic model.

\section{Testing the incremental model for bacteria}
We analyzed the results of video-microscopy experiments performed by Stewart et al. \cite{stewart} and Wang, Robert et al. \cite{jun}. Stewart et al. followed cells of \emph{E. coli}, strain MG1655 growing into microcolonies on LB-agarose pads at 30$^o$ C (25 mins doubling time), with a 2 minutes temporal resolution. Stewart et al. reconstructed cell lineages and measured the length of each cell in the microcolony at each time step. In the data from Wang, Robert et al., MG1655 \emph{E. coli} cells were grown in LB medium at 37$^o$ C in microchannels and the length of the cells was measured every minute (doubling time 20 mins). Due to the microchannels structure, at each division only the old-pole daughter cell is followed. From each dataset we extracted the results of several independent experiments (respectively 8 and 4 experiments). Each experiment of the agarose dataset corresponds to the growth of 6 microcolonies with up to 600 cells (the first 150 minutes of growth were discarded to ensure steady state growth) and each experiment of the $\mu$channels dataset to the growth of bacteria in a hundred microchannels for 40 generations (we kept only the first 50 generations of growth to avoid replicative aging effect and discarded the first 10 generations of those to ensure steady-state). Variations of cell width being negligible compared to variations in length we consider that length is equivalent to volume. Note that both datasets were generated by the analysis of fluorescent images (the bacteria constitutively express the Yellow Fluorescent Protein) using two different softwares.

\begin{figure}[h!]
A   \hspace{10 cm} B \\
\includegraphics[width=0.4\linewidth]{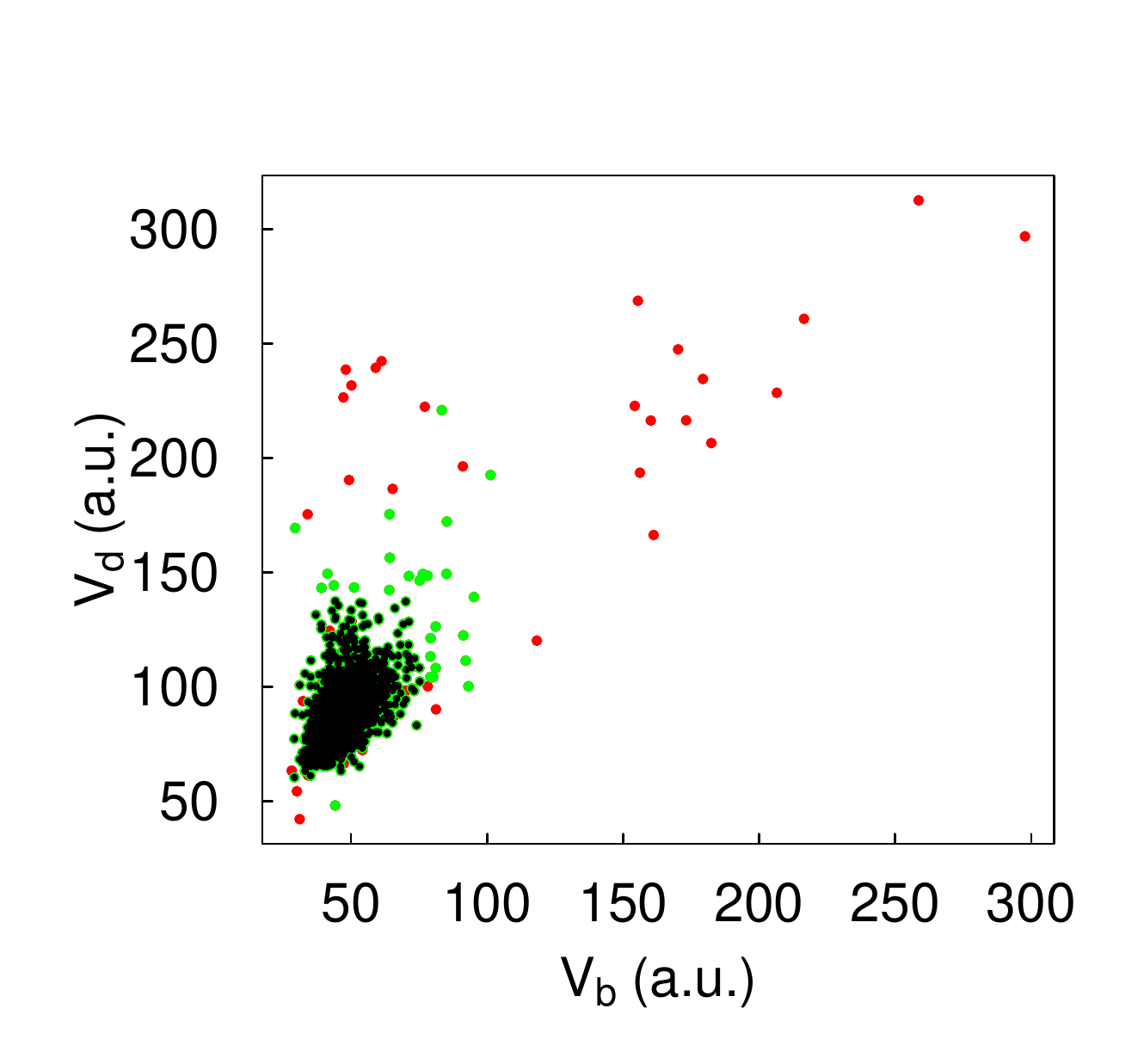} \hspace{2.5 cm}
\includegraphics[width=0.45\linewidth]{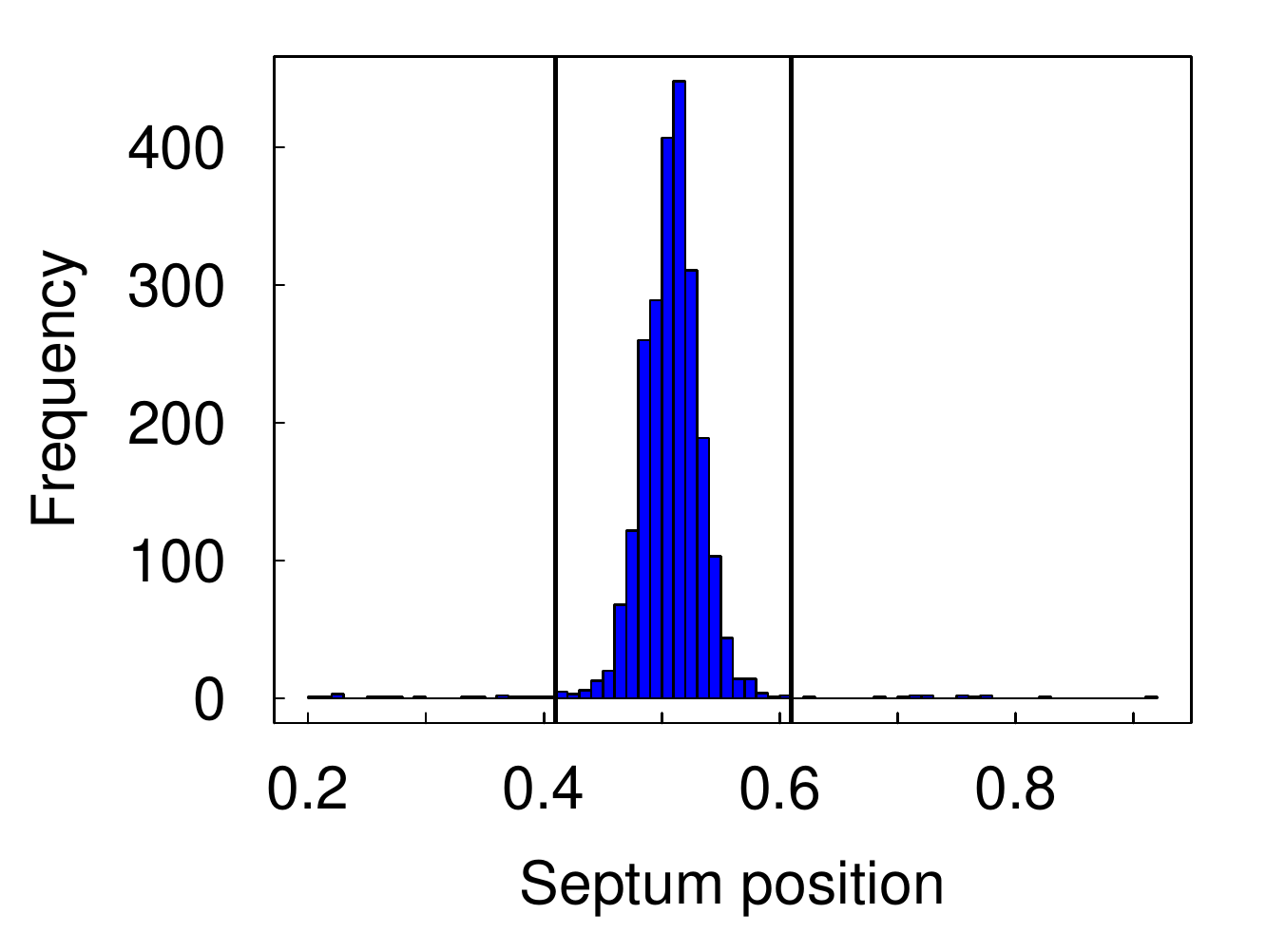}

\caption{\textbf{Analysis of the bacteria datasets, related to Figure 4.}
\textbf{A Size at division versus size at birth in an experiment of the microchannels dataset.} Red dots : cells that exhibit asymmetric divisions and their daughters; Green dots: outliers that can be removed by simple thresholding; black dots: data with all outliers removed.  \textbf{B Histogram of septum position.}  Asymmetric divisions are detected by a septum position above 0.6 or below 0.4 (vertical lines).}
\label{out}
\end{figure}

Both the agarose and microchannels dataset contain a few outliers (1-2\% of the cells) that influence strongly the calculation of the correlation coefficients (see Fig. S\ref{out}A). In both agarose and microfluidic experiments, some cells filament and exhibit an asymmetric division: the septum is positioned at the quarter of the cell instead of the middle. Most of the outliers in the data (for either size at birth, size at division or interdivision time) correspond to cells that exhibit such asymmetric division or their daughters (see red dots in Fig. S\ref{out}A). Once these cells are removed (Fig. S\ref{out}B), the few remaining outliers can be removed by filtering the particularly high or low values of interdivision time, size at birth and size at division (see green dots in Fig. S\ref{out}A). The thresholds are calculated as the quantiles of the lognormal (for sizes) or normal (for times) distributions at the level 1/[sample size] (i.e. the probability to have a value above the maximum threshold or below the minimum threshold is 1/[number of cells]). Importantly, in contrast to the outliers created by asymmetric division, these outliers are not very influential and the thresholding procedure does not change significantly the correlation coefficients (example in Fig. S\ref{out}A).

 \emph{E. coli} growth occurs uniformly and the division is nearly symmetric, with the asymmetry coefficients (defined here as the ratio of sizes of a given daughter cell to the cell before division) distributed narrowly and normally \cite{size2,marr,dekker2}. For this reason the derivations in section 5 is not valid for \emph{E. coli}; even though $\lambda T$ may significantly differ between growth media, the asymmetry ratio is always close to 0.5. Another important difference is the existence of multiple replication forks in bacteria - implying that several divisions may occur during the DNA replication process. Despite these important differences, predictions A and B of the main text are still intact, as is shown in Refs. \cite{amir_PRL_size, ho}.

Predictions A and B are tested in Fig 4A of the main text. It shows that the correlations between the size at birth and size at division are equivalent to those of Fig. 1C of the main text, and that the negative correlations between the size at birth and the interdivision time are equivalent to Fig. 2, without using any fitting parameters.

The validity of the incremental model for \emph{E. coli} can be further tested by comparing theory and experiments for the correlation coefficients: in Ref. \cite{amir_PRL_size} the correlation coefficient between size at birth and size at division is shown to be 0.5 for perfectly symmetric division, and the correlation coefficient between size at birth and interdivision time is shown to be $-0.5$. These values are consistent with those measured in slow growth conditions \cite{correlation} and more recently in fast growth conditions \cite{size2}.

\end{document}